\documentclass[12pt]{article}
\DeclareUnicodeCharacter{2082}{$_2$}
\usepackage[utf8]{inputenc}
\usepackage{amssymb}
\usepackage{amsmath}
\usepackage{array}
\usepackage{graphicx,hyperref}
\usepackage[export]{adjustbox}
\usepackage{xcolor}
\usepackage{subcaption}
\usepackage{fancyhdr}
\title{\textbf{Dislocation Patterning as a Mechanism for \\ Flat Band Formation}}

\author{
    \textbf{Aziz Fall} \\
    \small Department of Mechanical Engineering, Carnegie Mellon University \\[-0.5em]
    \small \texttt{afall@andrew.cmu.edu} \\[1em]
    \textbf{Kaushik Dayal} \\
    \small Department of Civil and Environmental Engineering, Carnegie Mellon University \\[-0.5em]
    \small Center for Nonlinear Analysis, Department of Mathematical Sciences, Carnegie Mellon University \\[-0.5em]
    \small Department of Mechanical Engineering, Carnegie Mellon University \\[-0.5em]
    \small Pittsburgh Quantum Institute, University of Pittsburgh
}

\begin{document}
\date{}
\maketitle

\thispagestyle{fancy}

\begin{abstract}
    We compute the second-order correction to the electronic dispersion relation of a free electron gas interacting with an effective electron-dislocation potential, derived from a modern quantized theory of dislocations. Our results demonstrate that dislocation patterning induces anisotropic flat bands in the electronic dispersion under specific strain fields and directions, referred to as ``magic'' parameters. These flat bands acquire non-zero curvature as the strain or direction deviates from these magic parameters.    
\end{abstract}

\section{Introduction}

Materials with electronic flat bands near their Fermi energy exhibit intriguing electronic properties due to the divergence of the effective mass and the enhanced electron-electron interactions relative to their kinetic energy \cite{Kravchenko2017}. Flat bands have been associated with phenomena such as superconductivity, superfluid transport, fractional quantum Hall effects, and spin liquids \cite{tovmasyan2016,hu2023,chan2022,roy2019,herzog-arbeitman2022,wang2012}. Recent research has highlighted the role of non-uniform strain as a key factor in the formation of flat bands \cite{andrade2023, meng2024, bi2019, kariyado2023, utama2019, yang2022,wan2023,mesple2021}, with moiré materials providing notable examples.

In this study, we investigate the influence of dislocation strain fields on electronic dispersion relations and their potential role in flat band formation. Dislocations are ubiquitous in crystalline materials and play a significant role in determining their functional properties, including electrical and thermal transport, optical behavior, magnetic ordering, and superconductivity \cite{li2019quantized, fogel2001, li2017tailoring, podolyak2019, breio2024, azhar2022}. Despite their prevalence, the effect of dislocation strain fields on electronic dispersion and flat band phenomena remains an area requiring further exploration.

While there have been numerous studies successfully demonstrating strain-mediated band engineering---showing, for example, that external mechanical strain can modify the band gap of materials~\cite{Boland2024}, enhance carrier mobilities in transition metal dichalcogenides~\cite{Datye2022}, and even induce phase transitions in nickelate films~\cite{Cui2024}---a particularly recent focus has been on the ability of strain to induce and influence flat bands.
Recent work has shown that flat bands can emerge at non-magic angles in twisted bilayer graphene due to in-plane biaxial strain~\cite{Li2025}, and that strain can render flat bands anisotropic in Kagome lattices~\cite{Xu2024}.

What distinguishes the present study from previous work on strain-mediated flat bands in 2D materials---such as the analytical Bistritzer-MacDonald model of moiré flat bands~\cite{Bistritzer2011}---is not only that we investigate a different physical system, but also that we introduce a parameter in our modified dislocation potential to help tune the strain field. Our model allows for statistical variation of this parameter, and consequently, statistical variation in the strain field of the system.

Building on prior work that quantized dislocation displacement fields to derive an effective electron-dislocation interaction \cite{li2018}, we introduce an additional factor that tunes the dislocation strain field of the interaction. This modification allows us to study how different dislocation strain fields alter the electronic dispersion of a free electron gas. 

While there exist traditional frameworks for studying the influence of dislocation dynamics on solid-state materials and their electronic structure---such as continuum dislocation dynamics using dislocation velocity tensor fields~\cite{ElAzab2020}, phase-field modeling involving the evolution of a continuous order parameter representing dislocation density~\cite{Zheng2018,beyerlein2016understanding,koslowski2002phase,peng20203d}, discrete dislocation dynamics~\cite{Zbib2012}, molecular dynamics simulations~\cite{Yin2021}, and modeling dislocations as lines of point defects in density functional theory (DFT) calculations~\cite{Park2016}---these approaches are generally suited for capturing large-scale or mean-field changes in the dislocation. However, the frameworks employed in these methods make it difficult to study the impact of dislocations on quantum many-body systems and their influence on functional properties such as superconductivity and magnetism~\cite{li2018}. Moreover, simulating dislocations in complex systems like high-entropy alloys or curved dislocations using DFT requires modeling millions of atoms~\cite{Mismetti2024}, making such calculations computationally expensive.

In contrast, employing an effective field theory based on the recently developed concept of quantized dislocations (dislons) provides a powerful alternative. This approach enables the incorporation of many-body effects and techniques, allowing the study of dislocation-induced modifications to electronic properties without relying on traditional mean-field DFT calculations. It also offers a unified framework for treating both vibrational and static strain effects analytically, without requiring empirical fitting parameters to model the dislocation potential~\cite{li2018}.

In this work, we adopt the dislocation quantization scheme developed by Li \textit{et al.}~\cite{li2018} to obtain an effective electron--dislocation interaction. Although the quantization approach they use to model quantum dislocation dynamics is not unique, they are other quantization schemes that treat the dislocation as a string field rather than the scalar field adopted here \cite{Lund2019, Lin2022}, our chosen framework has the advantage of yielding a Lagrangian for the electron-dislocation interaction that is easily integrable over the dislon fields, resulting in an effective theory solely in terms of electron field operators. One limitation of the scalar-field-based dislon quantization adopted in this work is the absence of longitudinal dislon modes, but this is outweighed by the advantage of an analytically simpler Lagrangian.

The paper is organized as follows: in Section \ref{sec:theory_overview}, we provide a summary of the theoretical framework underlying our approach. Section \ref{sec:strain_tuning} presents the derivation of the dislocation strain field tuning factors. Sections \ref{sec:Numerical_integration} and \ref{sec:Numerical_Dispersion} detail our numerical techniques. Results are discussed in Section \ref{sec:Results}, and we conclude with a summary of findings and implications in Section \ref{sec:conclusion}.

\section{Theoretical Overview}
\label{sec:theory_overview}
We begin with a concise overview of the theory of quantized dislocations developed by Li et al \cite{li2018}. Similar to phonons, dislocations represent displacements of atoms within a lattice from a reference equilibrium configuration. However, unlike phonons, dislocation displacements in the continuum limit are governed by a topological constraint given by  
\begin{equation} 
\oint_{D} d\mathbf{u} = -\mathbf{b}, 
\label{constraint} 
\end{equation} 
where $\mathbf{u}$ denotes the displacement field, and $\mathbf{b}$ is the Burgers vector. A dislon is defined as the quantized excitation along the dislocation line, where the displacement field of the zeroth-mode excitation satisfies the constraint in Equation \ref{constraint}. 

If we orient the dislocation line along the \( z \)-axis, a general mode expansion of the displacement field can be expressed as 
\begin{equation}
    \mathbf{u}(\mathbf{R}) = \frac{1}{L^2} \sum_{\mathbf{k}} e^{i \mathbf{k} \cdot \mathbf{R}} e^{-i \mathbf{s} \cdot \mathbf{r_{0}}} \mathbf{U_{k}},
\end{equation}
where \(\mathbf{k} = (k_{x}, k_{y}, k_{z})\), \(\mathbf{s} = (k_{x}, k_{y}, 0)\), \( L \) is the crystal size, and \(\mathbf{r_{0}}\) denotes the coordinate of the dislocation core. The classical displacement field is obtained by taking the limit as \( k_{z} \to 0 \) in the mode expansion. This static field can be written as 
\begin{equation}
  \mathbf{u}(\mathbf{R}) |_{stat} = \frac{1}{L^2} \sum_{\mathbf{s}} e^{i \mathbf{s} \cdot \mathbf{r}} e^{-i \mathbf{s} \cdot \mathbf{r_{0}}} \mathbf{F(s)},
  \label{static_mode_expansion}
\end{equation}
where \(\mathbf{F(s)}\) is defined as \(\lim_{k_{z} \to 0} \mathbf{U_{k}} = \mathbf{F(s)}\) and is chosen so that the static displacement field adheres to the constraint in Equation \ref{constraint}. Consequently, we can express \(\mathbf{U_{k}}\) as 
\begin{equation}
    \mathbf{U_{k}} \equiv u_{\mathbf{k}} \mathbf{F}(\mathbf{k}),
\end{equation}
where \( u_{\mathbf{k}} \) is a scalar function and \( \mathbf{F}(\mathbf{k}) \) serves as a polarization vector. Quantizing the dislocation displacement field thus reduces to quantizing a scalar field, with a constraint that \( u_{\mathbf{k}} \) satisfies
\begin{equation}
      \lim_{ k_{z} \to 0  } u_{\mathbf{k}} = 1.
      \label{boundary_condition}
\end{equation}

To proceed with quantization, we express the classical dislocation Hamiltonian as 
\begin{equation}
    H_{D} = \frac{\rho}{2} \int \sum_{i = 1}^{3} \dot{\mathbf{u}_{i}^2}(\mathbf{R}) \, d^{3}\mathbf{R} + \frac{1}{2} \int c_{ijkl} u_{ij} u_{kl} \, d^{3}\mathbf{R},
\end{equation}
where \( \rho \) is the mass density, \( c_{ijkl} \) is the stiffness tensor for an isotropic material, defined by \( c_{ijkl} \equiv \lambda \delta_{ij} \delta_{kl} + \mu (\delta_{ik} \delta_{jl} + \delta_{il} \delta_{jk}) \) with \( \lambda \) and \( \mu \) as the Lamé parameters, and \( u_{ij} \) is the strain tensor. 

Defining \( u_{\mathbf{k}} \) as the canonical coordinate and \( \mathbf{\mathcal{L}} \) as the corresponding Lagrangian in reciprocal space, we can then define the conjugate momentum as 
\begin{equation}
    p_{\mathbf{k}} = \frac{\partial \mathbf{\mathcal{L}}}{\partial \dot{u_{\mathbf{k}}}}.
\end{equation}
Thus, the Hamiltonian in first-quantized form can be written as 
\begin{equation}
    H_{D} = \frac{1}{2} \sum_{\mathbf{k}} \frac{ p_{\mathbf{k}} p_{\mathbf{-k}}}{ m_{\mathbf{k}}} + \frac{1}{2L} \sum_{ \mathbf{k} } W_{\mathbf{k}} u_{\mathbf{k}} u_{ \mathbf{-k} },
\end{equation}
where \( m_{ \mathbf{k}} \equiv \rho|\mathbf{F}(\mathbf{k})|^{2}/L \) and \( W_{\mathbf{k}} \equiv (\lambda + \mu)[ \mathbf{k} \cdot \ \mathbf{F}(\mathbf{k})]^{2} + \mu k^2 |\mathbf{F}(\mathbf{k})|^{2} \). The canonical creation and annihilation operators are defined by the equations 
\begin{equation}
    u_{\mathbf{k}} = Z_{\mathbf{k}} [ a_{\mathbf{k}} + a_{-\mathbf{k}}^{\dagger}]
\end{equation}
and 
\begin{equation}
    p_{\mathbf{k}} = \frac{i \hbar}{2Z_{\mathbf{k}}} [ a_{\mathbf{k}}^{\dagger} - a_{-\mathbf{k}}],
\end{equation}
with \( Z_{\mathbf{k}} \equiv \sqrt{\frac{\hbar}{2 m_{\mathbf{k}} \Omega_{\mathbf{k}} } } \) and \( \Omega_{\mathbf{k}} \equiv \sqrt{\frac{W_{\mathbf{k}}}{m_{\mathbf{k}}L}} \). Due to the constraint in Equation \ref{boundary_condition}, \( a_{\mathbf{k}} \) and \( a_{\mathbf{k}}^{\dagger} \) do not follow the canonical bosonic commutation relations \( [a_{\mathbf{k}}, a_{\mathbf{k'}}^{\dagger}] = \delta_{\mathbf{k}\mathbf{k'}} \). Enforcing this relation leads to mathematical inconsistencies, so an alternative quantization scheme is used, where the operators satisfy 
\begin{equation}
    [a_{\mathbf{k}}, a_{\mathbf{k'}}^{\dagger}] = \delta_{\mathbf{k} \mathbf{k'}} \, \operatorname{sgn}(\mathbf{k}),
\end{equation}
where \( \operatorname{sgn}(\mathbf{k}) \) is defined as 
\begin{align}
    & \operatorname{sgn}(\mathbf{k}) = \begin{cases} 
                    +1, & \text{if} \ k_{x} > 0 \\
                    -1, & \text{if} \ k_{x} < 0 \\ 
                    \operatorname{sgn}(k_{y},k_{z}), & \text{if} \ k_{x} = 0
                  \end{cases} \\
    & \operatorname{sgn}(k_{y},k_{z}) = \begin{cases} 
                    +1, & \text{if} \ k_{y} > 0 \\
                    -1, & \text{if} \ k_{y} < 0 \\
                    \operatorname{sgn}(k_{z}), & \text{if} \ k_{y} = 0
                  \end{cases} \\
    & \operatorname{sgn}(k_{z}) = \begin{cases} 
                    +1, & \text{if} \ k_{z} > 0 \\
                    -1, & \text{if} \ k_{z} < 0 \\
                    0, & \text{if} \ k_{z} = 0 
                  \end{cases}
\end{align}
We can now define a pair of new operators, each of which satisfies the canonical bosonic commutation relations, given by 
\begin{align}
    d_{\mathbf{k}} & = \frac{1}{\sqrt{2}}(a_{\mathbf{k}1} + a_{\mathbf{k}2})
    \\
    f_{\mathbf{k}} & = \frac{1}{\sqrt{2}}(-a_{\mathbf{k}1} + a_{\mathbf{k}2})
\end{align}
where $a_{\mathbf{k}1}$ and $a_{\mathbf{k}2}$ are defined according to the following relations: 
\begin{equation}
    a_{\mathbf{k}1} = a_{\mathbf{k}} \ \text{if} \ \operatorname{sgn} (\mathbf{k}) > 0 \quad ; \quad
     a_{-\mathbf{k}2}^{\dagger} = a_{\mathbf{k}} \ \text{if} \ \operatorname{sgn} (\mathbf{k}) < 0
\end{equation}
The second quantized Dislon Hamiltonian can then be written as 
\begin{equation}
    H_{D} = \sum_{\operatorname{sgn}(\mathbf{k}) > 0} \hbar \Omega_{\mathbf{k}} \left( d_{\mathbf{k}}^{\dagger} d_{\mathbf{k}} + \frac{1}{2}  \right)  + \sum_{\operatorname{sgn}(\mathbf{k}) > 0} \hbar \Omega_{\mathbf{k}} \left( f_{\mathbf{k}}^{\dagger} f_{\mathbf{k}} + \frac{1}{2}  \right)
\end{equation}
where the constraint given by Equation \ref{boundary_condition} translates to the following condition on the d-field: 
\begin{equation}
    \lim_{k_{z} \to 0} d_{\mathbf{k}} = \lim_{k_{z} \to 0} \sqrt{\frac{m_{\mathbf{k}} \Omega_{\mathbf{k}}}{\hbar}} 
    \label{dfield_constraint}
\end{equation}

We are now in a position to write down the electron-dislon interaction in terms of our field operators. We start by writing the electron-ion interaction as 
\begin{equation}
    H_{e-ion} = \int d^{3}\mathbf{R} \rho_{e}(\mathbf{R}) \sum_{j = 1}^{N} \nabla_{\mathbf{R}} V_{ei}(\mathbf{R} - \mathbf{R}_{j}^{0}) \cdot \mathbf{u}(\mathbf{R}_{j}^{0})
    \label{e-ion}
\end{equation}
where $\rho_{e}(\mathbf{R})$ is the electron charge density operator, $V_{ei}(\mathbf{R})$ is the electron-ion Coulomb potential, $\mathbf{u}(\mathbf{R} )$ is the dislocation displacement field, and $N$ is the number of ions in the material. In second quantized form, this turns into 
\begin{equation}
    H_{e-dis} = \sum_{\substack{\mathbf{k}^{'} \sigma \\ \operatorname{sgn}(\mathbf{k}) > 0}} \sqrt{2} g_{\mathbf{k}} e^{-i\mathbf{s} \cdotp \mathbf{r}_{0}} c_{\mathbf{k}^{'} + \mathbf{k} \sigma}^{\dagger}c_{\mathbf{k}^{'} \sigma} d_{\mathbf{k}} + \sum_{\substack{\mathbf{k}^{'} \sigma \\ \operatorname{sgn}(\mathbf{k}) > 0}} \sqrt{2} g_{\mathbf{k}}^{*} e^{+i\mathbf{s} \cdotp \mathbf{r}_{0}} c_{\mathbf{k}^{'} - \mathbf{k} \sigma}^{\dagger}c_{\mathbf{k}^{'} \sigma} d_{\mathbf{k}}^{\dagger}
\end{equation}
where $g_{\mathbf{k}} \equiv \frac{eN}{L^5} V_{\mathbf{k}} [i\mathbf{k} \cdot \mathbf{F}(\mathbf{k})] \sqrt{ \frac{ \hbar }{ 2 m_{\mathbf{k}} \Omega_{\mathbf{k}}}}$ and  $ V_{\mathbf{k}} = \frac{4\pi Z e}{k^2 + k_{TF}^2}$, with $Z$ being the effective nuclear charge, $k_{TF}$ is the Thomas-Fermi screening wavevector, and $e$ is the electron charge. $c_{\mathbf{k}}^{\dagger}$ and $c_{\mathbf{k}}$ are the electron creation and annihilation operators. 

Now that we have the electron-dislon potential in second quantized form, we can use the path integral formulation to write an effective potential for the electron-dislon interaction in terms of the electron field/operators alone by integrating over the dislon fields. This is possible since the interaction is linear in the d-field. The path integral formulation is also convenient in that the constraint of the d-field given by Equation \ref{dfield_constraint} can be easily incorporated. After integrating over the dislon degrees of freedom, the effective Hamiltonian becomes
\begin{equation}
    H_{eff} = \sum_{\mathbf{k} \sigma} ( \epsilon_{\mathbf{k}} - \mu)c_{\mathbf{k} \sigma}^{\dagger}c_{\mathbf{k} \sigma}  + \sum_{\mathbf{k} \mathbf{p} \mathbf{q} \sigma \sigma^{'}} V_{eff}(\mathbf{k}) c_{\mathbf{k} + \mathbf{p} \sigma}^{\dagger} c_{\mathbf{k} - \mathbf{q} \sigma^{'}}^{\dagger} c_{\mathbf{q} \sigma^{'}} c_{\mathbf{p} \sigma} + \frac{1}{L^2} \sum_{\mathbf{k} \sigma} \sum_{\mathbf{s}} ( A_{\mathbf{s}} e^{-i\mathbf{s} \cdotp \mathbf{r}_{0}} c_{\mathbf{k} + \mathbf{s} \sigma}^{\dagger} + A_{\mathbf{s}}^{*} e^{i \mathbf{s} \cdotp \mathbf{r}_{0}} c_{\mathbf{k} - \mathbf{s} \sigma}^{\dagger}) c_{\mathbf{k} \sigma}
\end{equation}
where $V_{eff} = - \frac{g_{\mathbf{k}}^{*} g_{\mathbf{k}}}{\hbar \Omega_{\mathbf{k}}}$ and $A_{\mathbf{s}} = \frac{eN}{2 L^3} V_{\mathbf{s}} [i \mathbf{s} \cdot \mathbf{F}(\mathbf{s})]$. Notice that the third term represents electron scattering off a static dislocation. Another way of seeing this is by using the static dislon displacement field given by Equation \ref{static_mode_expansion} to write down the electron-ion interaction given by Equation \ref{e-ion}. The main focus of this research paper is to use the classical electron-dislocation interaction given by the third term to study changes in the electronic dispersion relation of free carriers in materials subject to changes in dislocation strain fields.

\begin{figure}[ht!]
    \centering
    \begin{subfigure}[b]{0.49\textwidth}
        \includegraphics[width=\textwidth]{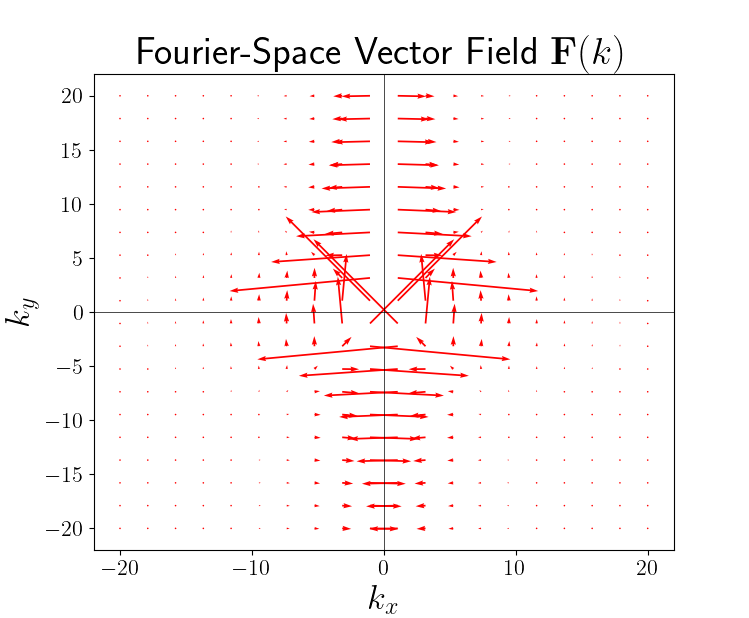}
        \caption{$\mathbf{F}(\mathbf{k})$ for an edge dislocation, as defined in Equation~\ref{F_k_eqn}.}
    \end{subfigure}
    \hfill
    \begin{subfigure}[b]{0.43\textwidth}
        \includegraphics[width=\textwidth]{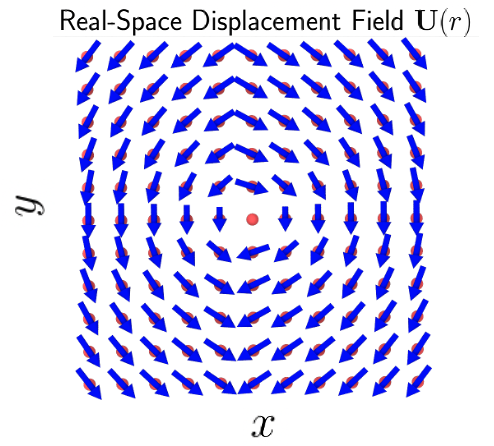}
        \caption{Real-space displacement field corresponding to $\mathbf{F}(\mathbf{k})$. Red dots mark the original atomic positions, and the center indicates the dislocation core.}
    \end{subfigure}
    \caption{Displacement fields in Fourier space (a) and real space (b) for an edge dislocation.}
    \label{fig:Disp_k_and_x}
\end{figure}

\section{Dislocation Patterning}
\label{sec:strain_tuning}
To analyze changes in the electronic dispersion relation due to dislocations, we must choose a form for \(\mathbf{F}(\mathbf{k})\) consistent with our dislocation potential. In this study, we focus on free carriers in materials affected by an array of edge dislocations. For an edge dislocation, \(\mathbf{F}(\mathbf{k})\) \cite{anderson2020theory, cai2016imperfections, stroh2021physics} takes the form [Fig \ref{fig:Disp_k_and_x}]:  
\begin{equation}
\label{F_k_eqn}
F_{z}(\mathbf{k}) = 0 \quad ; \quad F_{x}(\mathbf{k}) = \frac{b}{k_{x}s^2} \left( k_{y} - \frac{1}{1 - \nu} \frac{k_{x}^2 k_{y}}{s^2} \right)  \quad ; \quad F_{y}(\mathbf{k}) = \frac{b}{k_{x} s^2} \left( k_{x} - \frac{1}{1 - \nu} \frac{k_{y}^2 k_{x} }{s^2}  \right),
\end{equation}
where \(s^2 = k_{x}^2 + k_{y}^2\), \(b\) is the magnitude of the Burgers vector \(\mathbf{b} = (b,0,0)\), and \(\nu\) is the Poisson ratio. Consequently, the static dislocation potential \(A_{\mathbf{k}}\) becomes  
\begin{equation}
    A_{\mathbf{k}} = \frac{iN}{2 L^3} \frac{4 \pi Z}{s^2 + k_{TF}^2} \frac{2bk_{y}s^2 - 2\nu bk_{y}s^{2} - bk_{x}^2k_{y} - b k_{y}^{3}   }{s^{4}(1 - \nu)}.
\end{equation}
In the continuum limit (i.e., \(N \to \infty\) and \(L \to \infty\)), \(\frac{N}{L^3}\) is bounded and equals to the number density of ions per unit volume. We can now compute the first non-zero correction to the electronic self-energy for a random dislocation array (i.e., all the dislocation lines are along the \(z\)-axis, with randomly positioned cores) as done in \cite{li2018} using techniques from random impurity scattering \cite{rammer2004}. In the continuum limit, the averaged electron self-energy at zero temperature for an unpolarized electron gas is  
\begin{equation}
    \Sigma^{(2)}(\mathbf{p}, p_{z} ,E) =  \frac{\eta_{dis}}{ 2 \pi^2} \int d^{2} \mathbf{k} \frac{|A_{\mathbf{k}}|^{2}}{E - \frac{p_{z}}{2} - \frac{(\mathbf{k} + \mathbf{p})^{2}}{2} + i\epsilon},
    \label{div_selfE}
\end{equation}
where \(\mathbf{p} = (p_{x}, p_{y})\) and \(\eta_{dis}\) is the dislocation density, with the lattice spacing taken to zero. This integral has an infrared divergence as \(\mathbf{k} \to 0\) due to the divergence of \(A_{\mathbf{k}}\) at \(\mathbf{k} = 0\).

To eliminate this infrared divergence, we introduce an additonal factor that tunes the dislocation strain field, transforming \(A_{\mathbf{k}}\) as follows:
\begin{equation}
    A_{\mathbf{k}} \to A_{\mathbf{k}}(1 - e^{i\mathbf{k} \cdotp \mathbf{a}}) \equiv A_{dip}(\mathbf{k}),
    \label{dipole_transform}
\end{equation}
where the additional factor corresponds to transforming the potential to that of a dislocation dipole, with \(\mathbf{a}\) as the dipole separation vector. Thus, the additional factor controls the dislocation patterning of the dislocation array, and the vector \(\mathbf{a}\) controls its strain field. Additionally, we can average over all possible dipole separations \(\mathbf{a}\) when computing the electron propagator and self-energy. To achieve this, we revert to a finite crystal size \(L\) and then transition to the continuum limit.

\begin{figure}[ht!]
    \centering
    \begin{subfigure}{0.47\textwidth}
        \includegraphics[width=\textwidth]{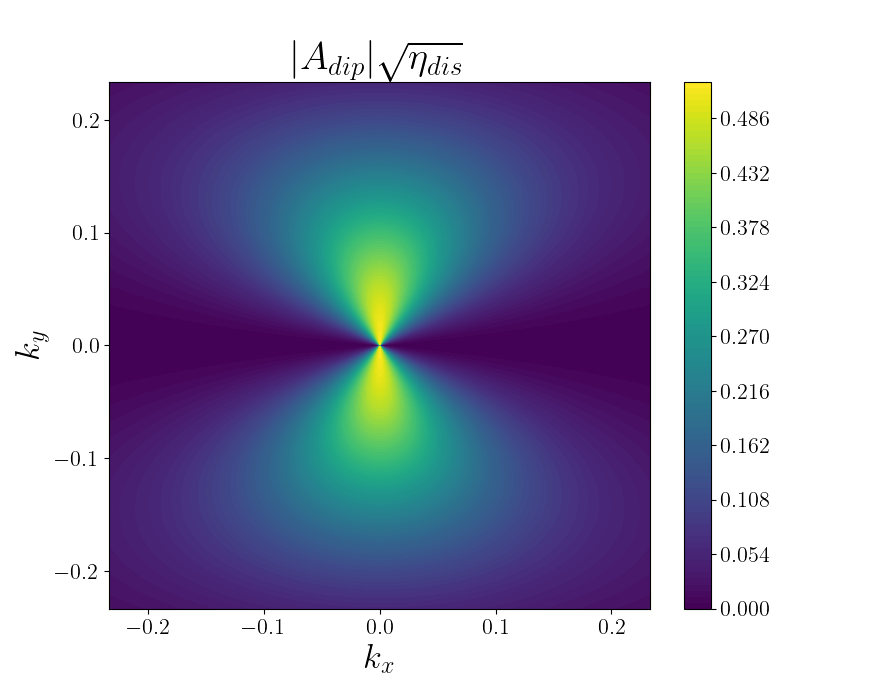}
        \caption{$A_{\mathrm{dip}}(\mathbf{k})$ weighted by the dislocation density.}
    \end{subfigure}
    \hfill
    \begin{subfigure}{0.47\textwidth}
        \includegraphics[width=\textwidth]{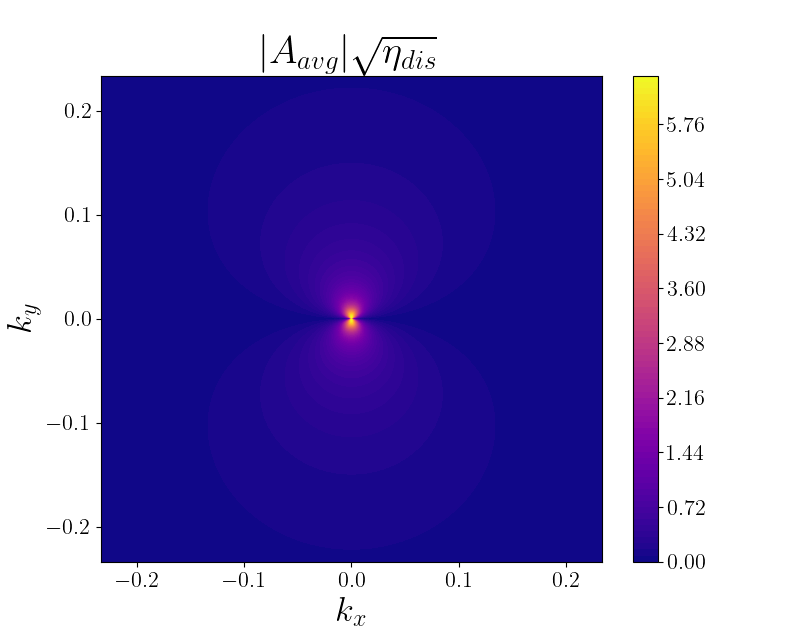}
        \caption{$A_{\mathrm{avg}}(\mathbf{k})$ weighted by the dislocation density.}
    \end{subfigure}
    \caption{Visual comparison of the potentials $A_{\mathrm{dip}}(\mathbf{k})$ and $A_{\mathrm{avg}}(\mathbf{k})$. Note that $A_{\mathrm{dip}}(\mathbf{k})$ varies over a broader region in $\mathbf{k}$-space compared to $A_{\mathrm{avg}}(\mathbf{k})$, even though both are plotted over the same domain and for the same set of physical parameters shared between the two potentials.}
    \label{fig:A_avg_and_dip}
\end{figure}

The Fourier transform of the full dislocation array potential, given a set distribution of dislocation core positions \(\mathbf{r}_{j}\) and dipole vectors \(\mathbf{a}_{j}\), is 
\begin{equation}
    \mathcal{F}\{A_{FULL}(\mathbf{r}_{1},\ldots,\mathbf{r}_{N_{dis}}; \mathbf{a}_{1},\ldots,\mathbf{a}_{N_{dis}})\} = A_{\mathbf{k}} \sum_{j = 1}^{N_{dis}} e^{-i\mathbf{k} \cdotp \mathbf{r}_{j}} \left( 1 - e^{i\mathbf{k} \cdotp \mathbf{a}_{j}}  \right),
\end{equation}
where \(N_{dis}\) is the number of dislocations. We assume the probability of finding a dislocation core at any given lattice site is i.i.d. with probability \(\frac{1}{L^2}\) (in the limit where lattice sites far exceed the number of dislocations). For the dipole separation vector \(\mathbf{a}_{j}\), we also assume the \(\mathbf{a}_{j}\) vectors associated with different dislocation cores to be i.i.d. The angular separations between paired dislocations relative to the x-axis are assumed to be uniformly distributed between \(0\) and \(2\pi\), while magnitudes \(a_j \equiv |\mathbf{a}_{j}|\) follow a nearest-neighbor Poisson distribution \cite{mathai1999}, where the mean point density is the dislocation density \(\eta_{dis}\). Thus, the probability of finding a dislocation core with associated \(\mathbf{a}_{j}\) is 
\begin{equation}
    P(\mathbf{a}_{j}) = P(\theta)P(a_{j}) = \frac{\eta_{dis} a_{j} e^{-\eta_{dis} \pi a_{j}^2}}{1 - e^{-\eta_{dis}\pi L^2}}.
\end{equation}
The averaged second-order electron propagator is 
\begin{equation}
\begin{aligned}
    \left\langle G^{(2)}(\mathbf{p}, \mathbf{p}', E) \right\rangle = & \, G_{0}(\mathbf{p}, E) \sum_{\mathbf{p}''} A_{\mathbf{p} - \mathbf{p}''} G_{0}(\mathbf{p}'', E) A_{\mathbf{p}'' - \mathbf{p}'} G_{0}(\mathbf{p}', E) \\
    & \times \left\langle \sum_{m, n = 1}^{N_{dis}} e^{-i(\mathbf{p} - \mathbf{p}'') \cdot \mathbf{r}_{m}} e^{-i(\mathbf{p}'' - \mathbf{p}') \cdot \mathbf{r}_{n}} 
    \times \left( 1 - e^{i(\mathbf{p} - \mathbf{p}'') \cdot \mathbf{a}_{m}} \right) \left( 1 - e^{i(\mathbf{p}'' - \mathbf{p}') \cdot \mathbf{a}_{n} } \right) \right\rangle,
\end{aligned}
\end{equation}
where \(G_{0}(\mathbf{p},E)\) is the free electron propagator. For simplicity, we absorbed some of the normalization factors into \(A_{\mathbf{k}}\). Thus, evaluating the second-order electron propagator is tantamount to evaluating the term in angular brackets. We compute this average for the case that \(\mathbf{r}_{m} = \mathbf{r}_{n} \equiv \mathbf{r}\) and \(\mathbf{a}_{m} = \mathbf{a}_{n} \equiv \mathbf{a}\). After averaging over dislocation cores, the expression in angular brackets reduces to  
\begin{equation}
    \eta_{dis} \delta_{\mathbf{p} \mathbf{p}'} \left \langle \left( e^{i\mathbf{a} \cdot (\mathbf{p} - \mathbf{p}'')} - 1 \right) \left(  e^{i\mathbf{a} \cdot (\mathbf{p}'' - \mathbf{p})} - 1   \right) \right \rangle.
    \label{avg1}
\end{equation} 
Taking the limit as $L \to \infty$  and then averaging over \(\mathbf{a}\), Equation \ref{avg1} simplifies to 
\begin{equation}
    2\eta_{dis} \delta_{\mathbf{p} \mathbf{p}'} \left( 1 - e^{-|\mathbf{k}|^{2}/(4\eta_{dis} \pi)} \right),
\end{equation}
where \(\mathbf{k} \equiv \mathbf{p} - \mathbf{p}''\). Thus, another method for eliminating the infrared divergence in Equation \ref{div_selfE} is to transform \(A_{\mathbf{k}}\) as
\begin{equation}
    A_{\mathbf{k}} \to A_{\mathbf{k}}\sqrt{2\left( 1 - e^{-|\mathbf{k}|^{2}/(4\eta_{dis} \pi)} \right)} \equiv A_{avg}(\mathbf{k}).
    \label{averaged_transform}
\end{equation}
However, this transformation does not provide a mechanism for tuning the strain field of the dislocation array. Therefore, we examine the electronic dispersion under both transformations. We will refer to the transformation governed by Equation \ref{dipole_transform} as the ``dipole transformation" and the one governed by Equation \ref{averaged_transform} as the ``averaged transformation" [Fig \ref{fig:A_avg_and_dip}].

Under the averaged transformation, the self-energy takes a functional form, $g(|\mathbf{p}|, p_{z}, E)$, where \( g \) is an arbitrary function. Notably, it remains invariant under the inversion \( \mathbf{p} \to -\mathbf{p} \). Interestingly, the self-energy under the dipole transformation also exhibits this invariance under \( \mathbf{p} \to -\mathbf{p} \), with a functional form given by \( h(|\mathbf{a}|, \mathbf{p} \cdot \mathbf{a}, |\mathbf{p}|, E, p_{z}) \), where \( h \) is also an arbitrary function.

To demonstrate this, we approximate \( A_{\mathbf{k}}(1 - e^{i\mathbf{k} \cdot \mathbf{a}}) \) as 
\[
A_{\mathbf{k}}(1 - e^{i\mathbf{k} \cdot \mathbf{a}}) \approx \frac{1}{k_{TF}^{2}} \frac{(1 - e^{i\mathbf{k} \cdot \mathbf{a}})}{|\mathbf{k}|}.
\]
Using Feynman parameters, we can then rewrite the self-energy integral as
\begin{equation}
\begin{aligned}
    \Sigma^{(2)}(\mathbf{p}, p_{z}, E) \approx & -\frac{4}{k_{TF}^4} \int_{0}^{1} dx \int d^{2}\mathbf{k} \frac{1}{\left( k^2 + \chi \right)^{2}} \\
    & + \frac{2}{k_{TF}^4} \int_{0}^{1} dx \, e^{-ix\mathbf{p} \cdot \mathbf{a}} \int d^{2}\mathbf{k} \frac{e^{i\mathbf{k} \cdot \mathbf{a}}}{\left( k^2 + \chi \right)^{2}} \\
    & + \frac{2}{k_{TF}^4} \int_{0}^{1} dx \, e^{ix\mathbf{p} \cdot \mathbf{a}} \int d^{2}\mathbf{k} \frac{e^{-i\mathbf{k} \cdot \mathbf{a}}}{\left( k^2 + \chi \right)^{2}}
\end{aligned}
\end{equation}
where $\chi \equiv -(x\mathbf{p})^{2} + x|\mathbf{p}|^{2} - 2x(E - \frac{p_{z}}{2}) -2xi\epsilon$, now it is easy to see the functional form of the self-energy integral and its invariance under the inversion $\mathbf{p} \to -\mathbf{p}$, recalling that $\mathbf{p} = (p_{x}, p_{y})$.

\section{Numerical Integration}
\label{sec:Numerical_integration}
To investigate changes in electronic dispersion caused by dislocation strain fields, we need to compute the self-energy. The integral in Equation \ref{div_selfE} under both dipole and averaged transformations is analytically challenging to evaluate. Furthermore, complex energy integration methods \cite{thiess2011} cannot be used to circumvent the pole slightly below the real axis, as energy integrations are not involved here. Therefore, we apply the Direct Computation Method (DCM) \cite{yuasa2012, deDoncker2004, deDoncker2004_} to compute the self-energy integral. The DCM algorithm involves two main steps:

\begin{enumerate}
    \item Compute the self-energy for a series of finite $\epsilon$ values (to avoid the pole along the real axis in Equation \ref{div_selfE}) using the sequence:
    \[
      \epsilon = \epsilon_{l} = \frac{\epsilon_{0}}{(A_{c})^{l}}, \quad l=0,1,\dots
    \]
    where $\epsilon_{0}$ and $A_{c}$ are constants, with $A_{c} > 1$.
    \item Extrapolate the integral value as $\epsilon_{l} \to 0$.
\end{enumerate}

In our calculations, we computed the self-energy from $l=0$ to $l=100$, using $\epsilon_{0} = 10$ and $A_{c} = 1.2$. We identified the inflection point of the series, where the integral’s concavity changes as a function of $l$, as the starting point for extrapolation. The series was terminated at the $l$ value where the integral began fluctuating significantly around the mean value over the preceding 13 iterations. We then applied Wynn's $\epsilon$-algorithm to extrapolate the integral’s value and compared this result by visually inspecting the integral’s behavior at the largest stable $l$ value before numerical instability set in, to see how close the values were without performing an extrapolation. At this order of perturbation theory, visual inspection of the series endpoint performed as well as, or better than, formal extrapolation. Thus, we computed the self-energy integrals by fixing $l$ at the largest stable value prior to fluctuation onset.

To validate this technique, we calculated the following representative integral, which resembles the second-order self-energy for random Coulomb scattering, though some factors specific to random Coulomb scattering have been omitted:
\begin{equation}
    M(E,\mathbf{p},k_{TF}) = \int d^{3}\mathbf{k} \frac{1}{E - \frac{(\mathbf{k} + \mathbf{p})^{2}}{2} + i\epsilon} \frac{1}{(|\mathbf{k}|^{2} + k_{TF}^{2})^{2} }
\end{equation}
for which an analytical solution exists (refer to Appendix \ref{appendix} for a more detailed calculation):
\begin{equation}
    M(E,\mathbf{p},k_{TF}) = \frac{-4\pi^{2}}{2k_{TF} \left( 2k_{TF} \sqrt{-2(E + i\epsilon)} -2(E + i\epsilon) + |\mathbf{p}|^{2} + k_{TF}^{2} \right) }
\end{equation} 
Using parameters $E = 0.3$, $k_{TF} = 0.2336$, and $\mathbf{p} = \frac{1}{2}\left(\frac{1}{\sqrt{3}},\frac{1}{\sqrt{3}},\frac{1}{\sqrt{3}}\right)$, the analytical result is $\mathcal{R}e\{M(E,\mathbf{p},k_{TF})\} = 114.385$ and $\mathcal{I}m\{M(E,\mathbf{p},k_{TF})\} = -140.117$, while our numerical result, obtained using $l=59$ through visual inspection, is $\mathcal{R}e\{M(E,\mathbf{p},k_{TF})\} = 114.409$ and $\mathcal{I}m\{M(E,\mathbf{p},k_{TF})\} = -139.118$. This close agreement demonstrates the DCM method’s validity for second-order self-energy calculations. Although we are confident in our second-order calculations, we note that the DCM Algorithm is not reliable in the context of our problem for higher order corrections, due to fluctuations of the self-energy integral before stable values was reached.

\section{Numerical Computation of Dispersion Relations}
\label{sec:Numerical_Dispersion}
In order to compute the electronic dispersion relation, we need to assign values to the parameters within our model. We use the following parameters expressed in atomic units, which are representative of Silicon (Table \ref{tab:parameters}):

\begin{table}[h!]
    \centering
    \begin{tabular}{|>{\centering\arraybackslash}m{2.5cm}|>{\centering\arraybackslash}m{3.5cm}|}
        \hline
        \textbf{Parameter} & \textbf{Value} \\
        \hline
        $b$ & $7.25571$ \\
        \hline
        $\nu$ & $0.275$ \\
        \hline
        $k_{\text{TF}}$ & $0.2336$ \\
        \hline
        $Z$ & $4.285$ \\
        \hline
        $\frac{N}{L^3}$ & $0.007403$ \\
        \hline
        $\eta_{\text{dis}}$ & $2.1 \times 10^{-6}$ \\
        \hline
        $l$ & $72$ \\
        \hline
        cutoff & $30 \times k_{TF}$ \\
        \hline
    \end{tabular}
    \caption{Model Parameters}
    \label{tab:parameters}
\end{table}

Here, $l$ represents the value used in the DCM algorithm for our self-energy integrations. The ``cutoff" parameter is employed because the momentum integrations are performed within a square of side length equal to the cutoff. This approach is necessary as numerical integration from negative infinity to positive infinity is impractical. We are free to choose any cutoff as long as the cutoff is chosen to be large enough such that the value of the integral remains invariant upon further increase. We saw that the self-energy integral for our chosen cutoff of $30$x$k_{TF}$ remained invariant upon increasing the cutoff any further.

The electronic dispersion relation is computed by finding the roots of the following equations:
\begin{align}
   f(E,\mathbf{k}, \mathbf{a}) & \equiv E - \epsilon_{\mathbf{k}} - \mathcal{R}e\{\Sigma_{\text{dip}}(\mathbf{k},E,\mathbf{a})\} = 0,
    \\
   f(E,\mathbf{k}) & \equiv E - \epsilon_{\mathbf{k}} - \mathcal{R}e\{\Sigma_{\text{Avg}}(\mathbf{k},E)\} = 0,
\end{align}
where $\epsilon_{\mathbf{k}} \equiv |\mathbf{k}|^{2}/2$. The first equation above represents the electron dispersion under the dipole transformation, with $\mathbf{a}$ being the dipole separation vector. The second equation represents the electronic dispersion under the averaged transformation.

\section{Results and Discussion}
\label{sec:Results}
We computed the electronic dispersion relations along several directions: $\hat{\alpha} = \left(1,0,0 \right)$, $\hat{\beta} = \left( \frac{1}{\sqrt{2}}, \frac{1}{\sqrt{2}},0 \right)$, $\hat{\gamma} = \left(0,1,0 \right)$, and $\hat{\Delta} = \left( 0, \frac{1}{\sqrt{2}}, \frac{1}{\sqrt{2}}\right)$. The computations employed the relation $\mathbf{a} = cb\hat{\mathbf{a}}$, where $\hat{\mathbf{a}} = \left(0,1 \right)$, \( b \) represents the magnitude of the Burgers vector, and \( c \) is an integer ranging from 1 to 4. 

\subsection{Averaged Transformation}

\begin{figure}[ht!]
    \centering
    \begin{subfigure}{0.47\textwidth}
        \includegraphics[width=\textwidth]{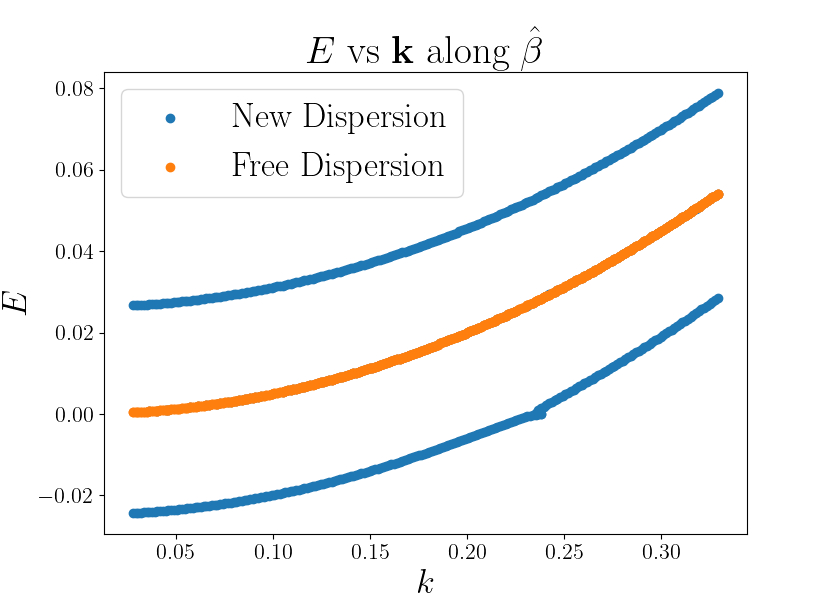}
        \caption{Dispersion relation plotted against the free dispersion curve.}
    \end{subfigure}
    \hfill
    \begin{subfigure}{0.47\textwidth}
        \includegraphics[width=\textwidth]{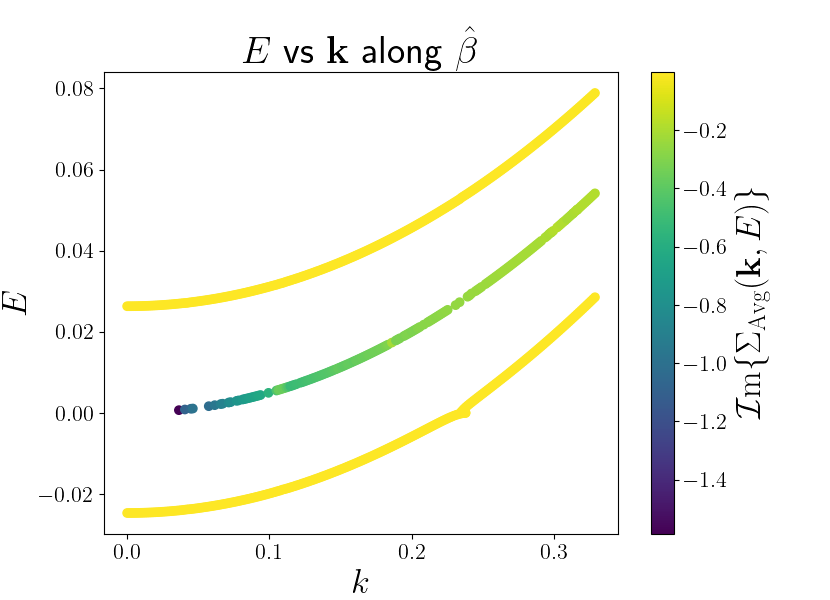}
        \caption{Colormap of the imaginary part of the self-energy.}
    \end{subfigure}
    \caption{Dispersion relation under the averaged transformation.}
    \label{fig:Avg_Trans}
\end{figure}

Figure \ref{fig:Avg_Trans} illustrates the dispersion relation under the averaged transformation, plotted against the free dispersion curve, \(\epsilon_{\mathbf{k}} = |\mathbf{k}|^{2}/2\). The interaction, as treated in second-order perturbation theory, yields three bands: one aligns with the free dispersion curve but with a reduced lifetime, while the other two correspond to constant energy shifts. These shifts primarily represent changes in the system's overall chemical potential. The low-energy band structure, corresponding to the free dispersion relation, appears highly unstable due to the reduced lifetime. The dispersion relation remains invariant under different directions as expected from the functional form of the self-energy derived previously. 

\begin{figure}[ht!]
    \centering
    \includegraphics[width=0.5\textwidth]{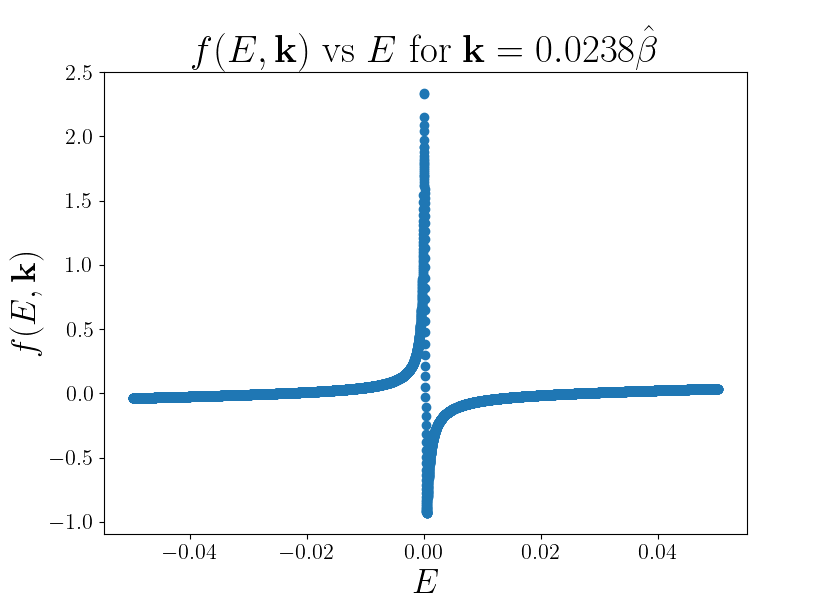}
    \caption{Behavior of \( f(E, \mathbf{k}) \) for \( \mathbf{k} = 0.0238\hat{\beta} \) under the averaged transformation.}
    \label{fig:f_e_k_avg}
\end{figure}

In part (b) of Figure \ref{fig:Avg_Trans}, missing data points for small \( |\mathbf{k}| \) arise from difficulties in resolving the energy root of the function \( f(E, \mathbf{k}) \) near the free energy root. This issue stems from sharp transitions in \( f(E, \mathbf{k}) \), as highlighted in Figure \ref{fig:f_e_k_avg}.

\begin{figure}[ht!]
    \centering
    \begin{subfigure}{0.47\textwidth}
        \includegraphics[width=\textwidth]{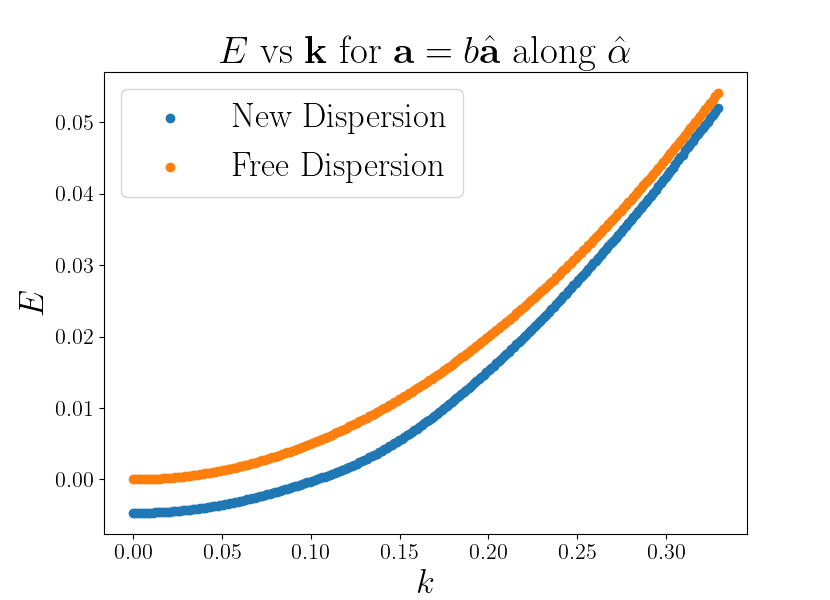}
        \caption{Dispersion relation for \( \mathbf{a} = b\hat{\mathbf{a}} \).}
    \end{subfigure}
    \hfill
    \begin{subfigure}{0.47\textwidth}
        \includegraphics[width=\textwidth]{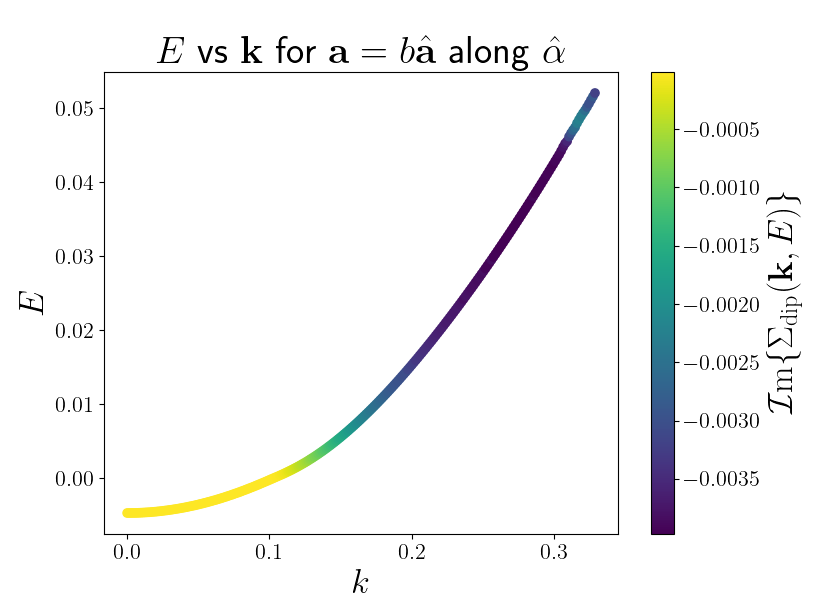}
        \caption{Colormap of the corresponding imaginary part.}
    \end{subfigure}
    \begin{subfigure}{0.47\textwidth}
        \includegraphics[width=\textwidth]{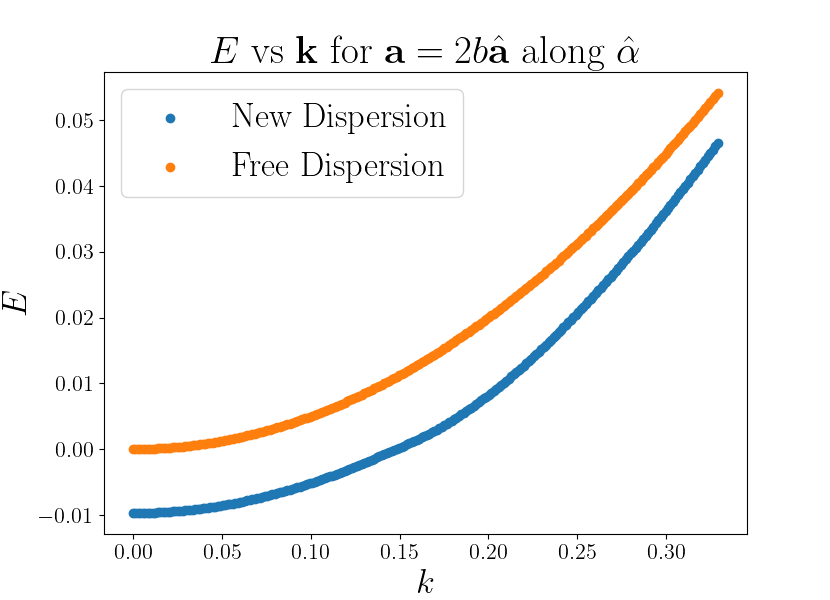}
        \caption{Dispersion relation for \( \mathbf{a} = 2b\hat{\mathbf{a}} \).}
    \end{subfigure}
    \hfill
    \begin{subfigure}{0.47\textwidth}
        \includegraphics[width=\textwidth]{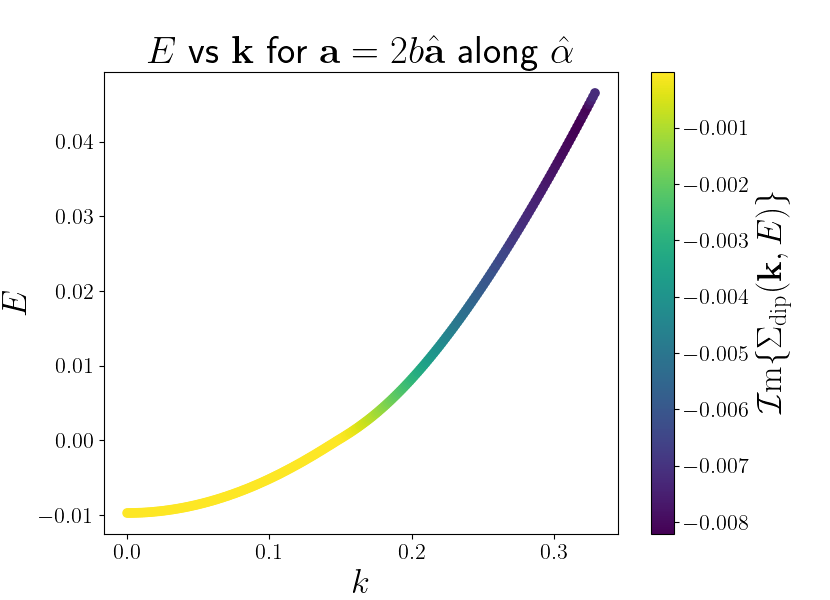}
        \caption{Colormap of the corresponding imaginary part.}
    \end{subfigure}
    \caption{Dispersion relations under the dipole transformation along \(\hat{\alpha}\).}
    \label{fig:Dipole_Alpha}
\end{figure}

\begin{figure}[ht!]
    \centering
    \begin{subfigure}{0.47\textwidth}
        \includegraphics[width=\textwidth]{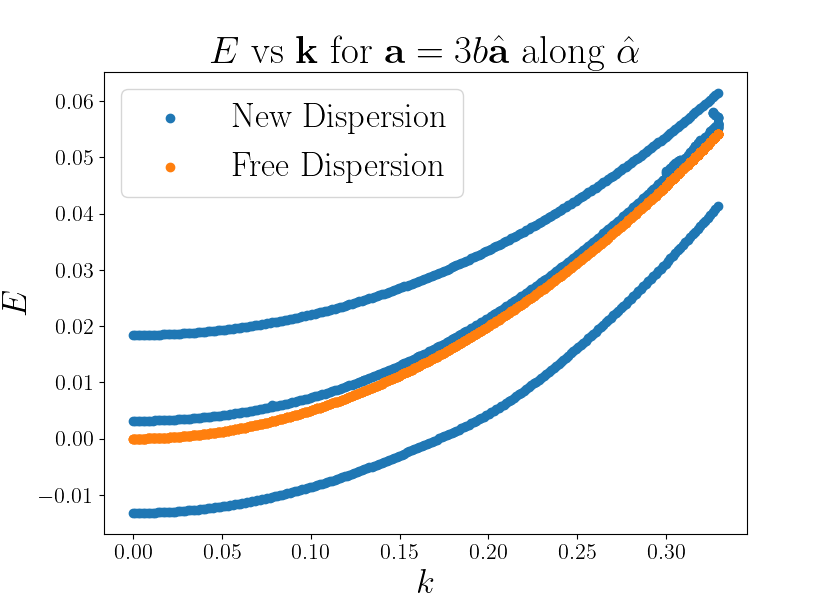}
        \caption{Dispersion relation for \( \mathbf{a} = 3b\hat{\mathbf{a}} \).}
    \end{subfigure}
    \hfill
    \begin{subfigure}{0.47\textwidth}
        \includegraphics[width=\textwidth]{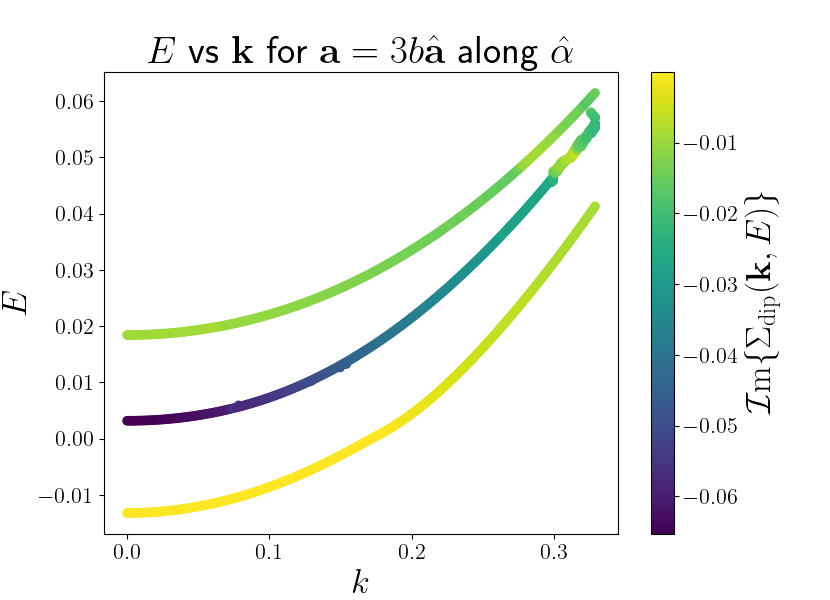}
        \caption{Colormap of the corresponding imaginary part.}
    \end{subfigure}
    \begin{subfigure}{0.47\textwidth}
        \includegraphics[width=\textwidth]{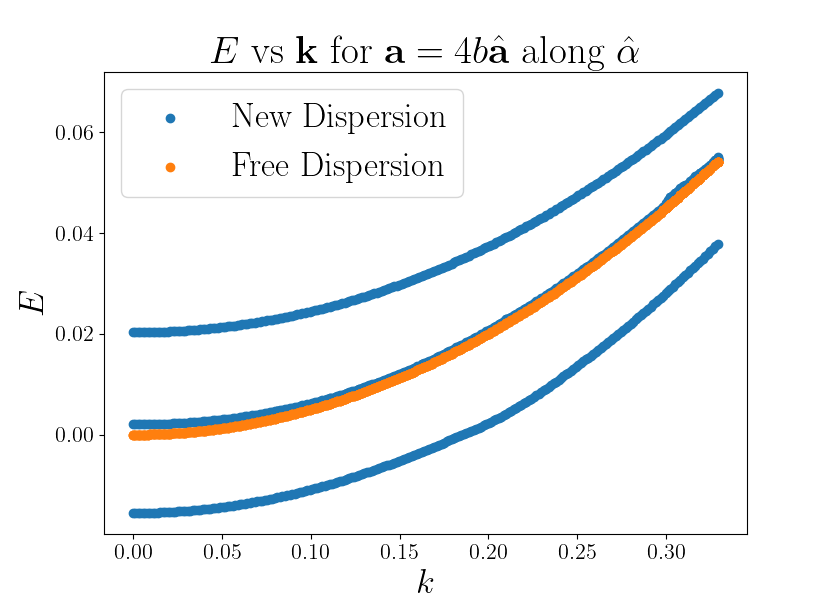}
        \caption{Dispersion relation for \( \mathbf{a} = 4b\hat{\mathbf{a}} \).}
    \end{subfigure}
    \hfill
    \begin{subfigure}{0.47\textwidth}
        \includegraphics[width=\textwidth]{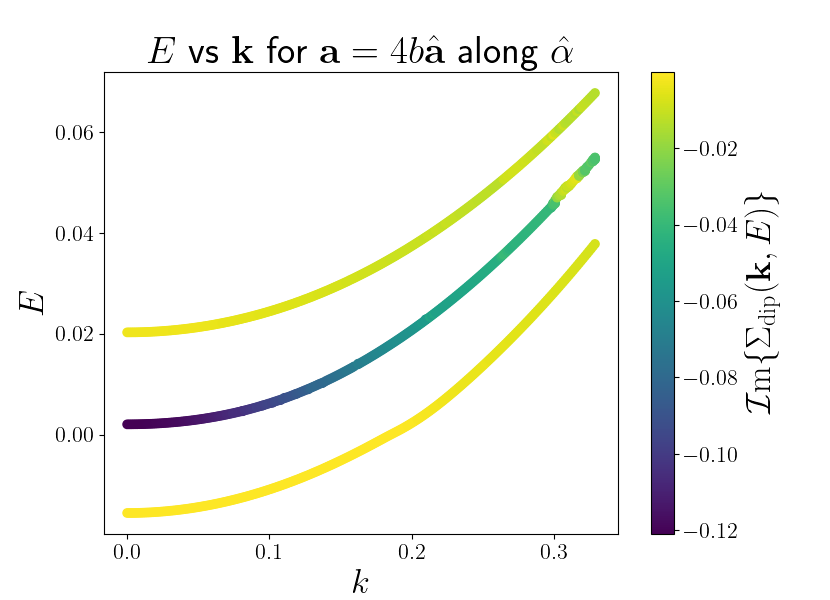}
        \caption{Colormap of the corresponding imaginary part.}
    \end{subfigure}
    \caption{Dispersion relations under the dipole transformation along \(\hat{\alpha}\) (continued).}
    \label{fig:Dipole_Alpha_Continued}
\end{figure}

\begin{figure}[ht!]
    \centering
    \begin{subfigure}{0.47\textwidth}
        \includegraphics[width=\textwidth]{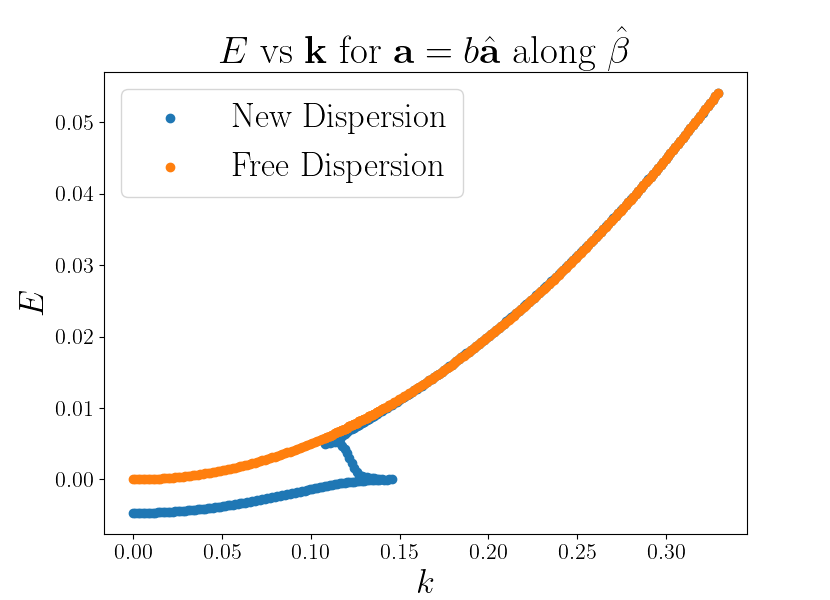}
        \caption{Dispersion relation for \( \mathbf{a} = b\hat{\mathbf{a}} \).}
    \end{subfigure}
    \hfill
    \begin{subfigure}{0.47\textwidth}
        \includegraphics[width=\textwidth]{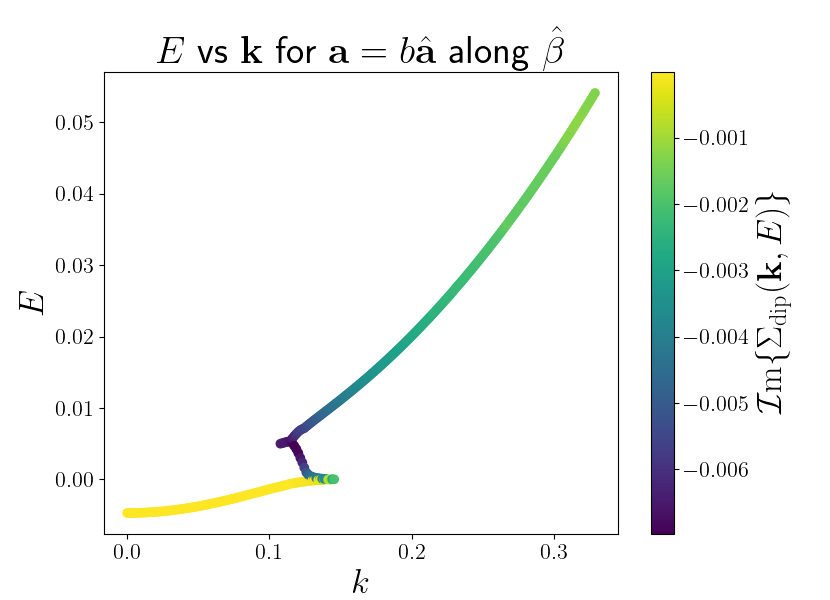}
        \caption{Colormap of the corresponding imaginary part.}
    \end{subfigure}
    \begin{subfigure}{0.47\textwidth}
        \includegraphics[width=\textwidth]{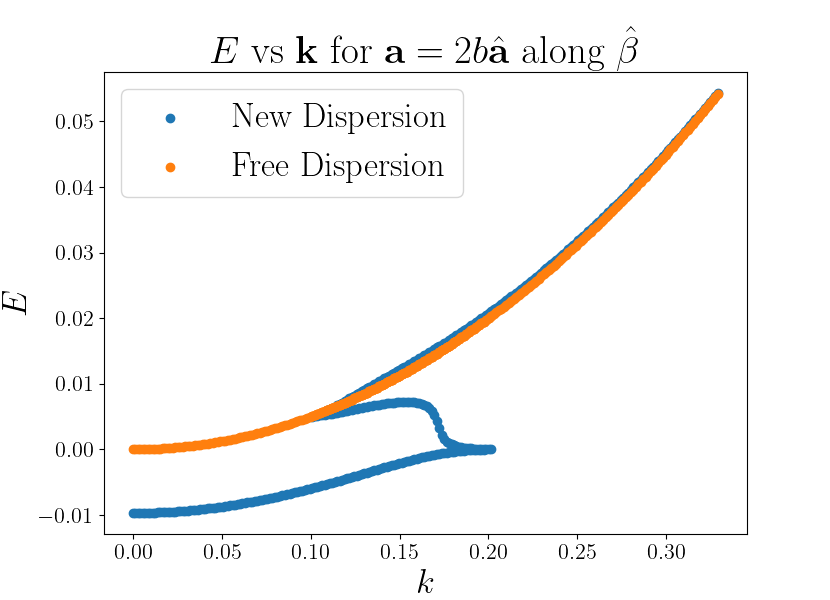}
        \caption{Dispersion relation for \( \mathbf{a} = 2b\hat{\mathbf{a}} \).}
    \end{subfigure}
    \hfill
    \begin{subfigure}{0.47\textwidth}
        \includegraphics[width=\textwidth]{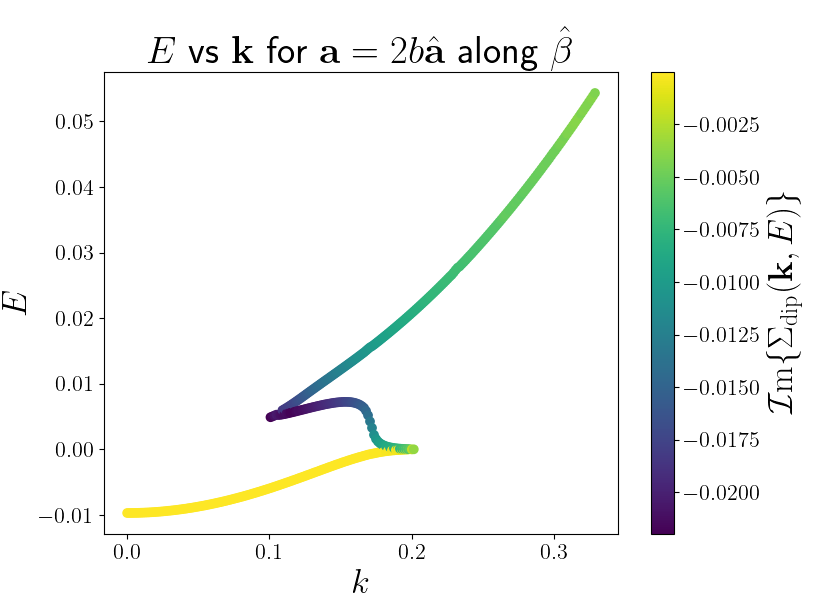}
        \caption{Colormap of the corresponding imaginary part.}
    \end{subfigure}
    \caption{Dispersion relations under the dipole transformation along \(\hat{\beta}\).}
    \label{fig:Dipole_Beta}
\end{figure}

\begin{figure}[ht!]
    \centering
    \begin{subfigure}{0.47\textwidth}
        \includegraphics[width=\textwidth]{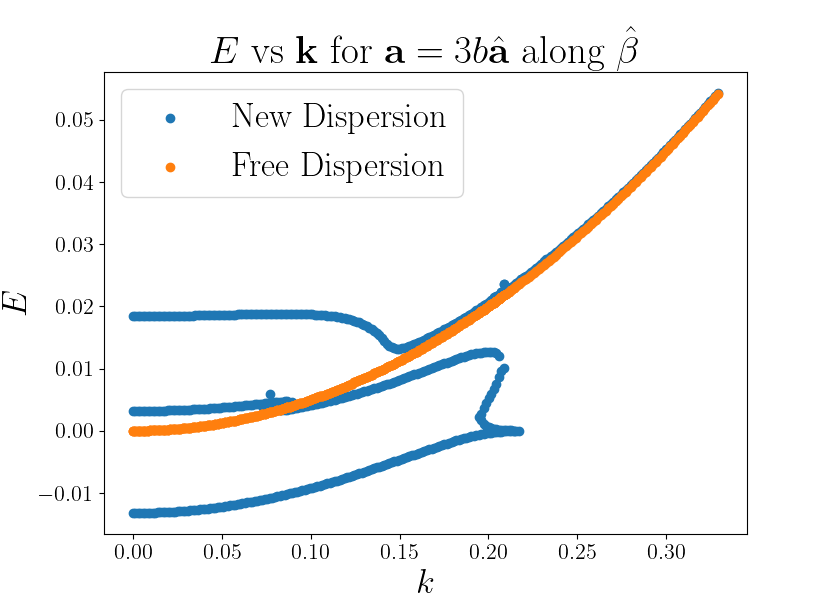}
        \caption{Dispersion relation for \( \mathbf{a} = 3b\hat{\mathbf{a}} \).}
    \end{subfigure}
    \hfill
    \begin{subfigure}{0.47\textwidth}
        \includegraphics[width=\textwidth]{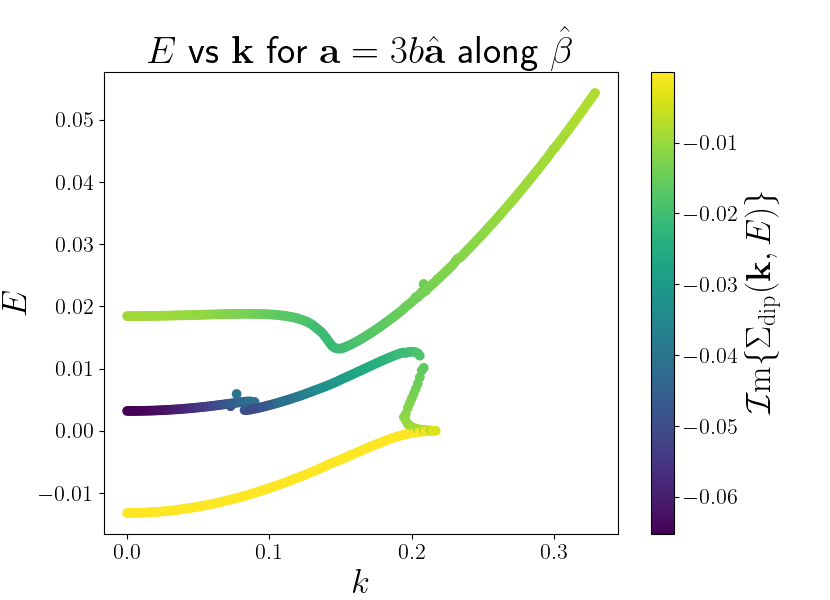}
        \caption{Colormap of the corresponding imaginary part.}
    \end{subfigure}
    \begin{subfigure}{0.47\textwidth}
        \includegraphics[width=\textwidth]{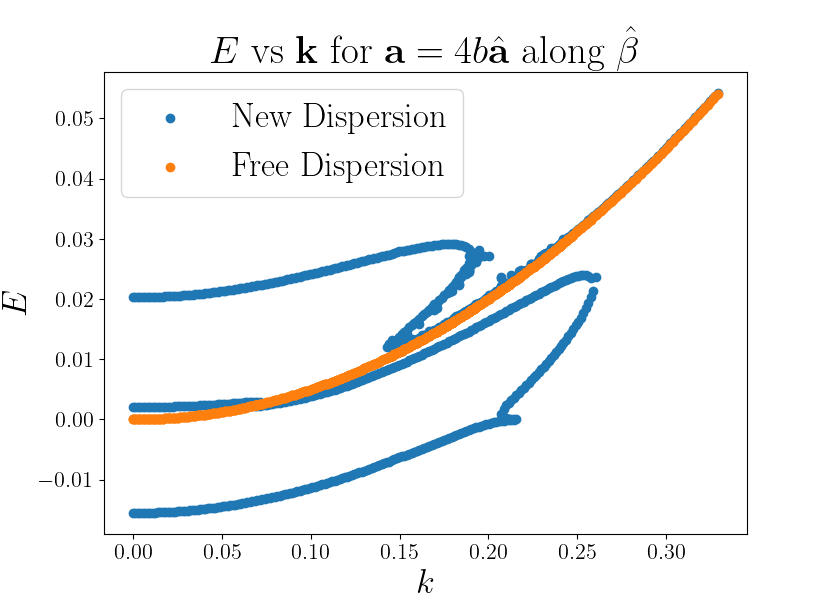}
        \caption{Dispersion relation for \( \mathbf{a} = 4b\hat{\mathbf{a}} \).}
    \end{subfigure}
    \hfill
    \begin{subfigure}{0.47\textwidth}
        \includegraphics[width=\textwidth]{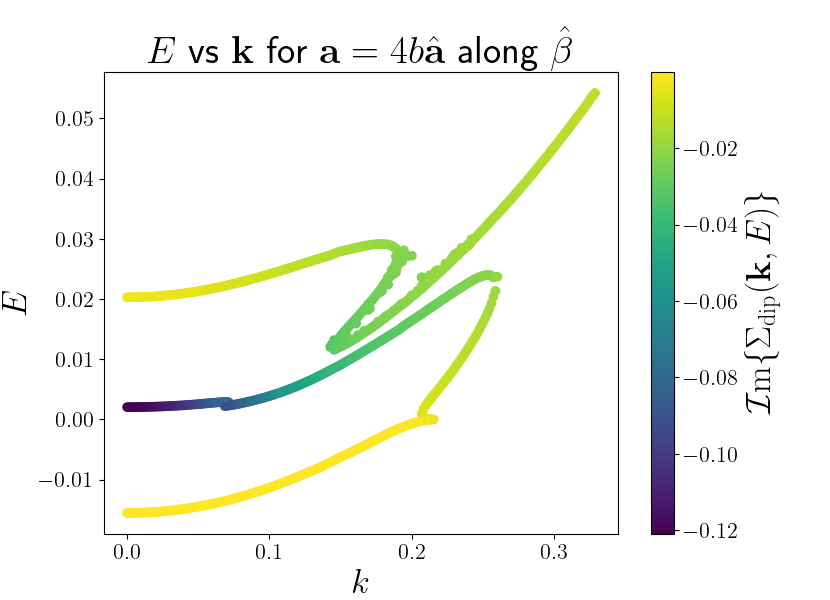}
        \caption{Colormap of the corresponding imaginary part.}
    \end{subfigure}
    \caption{Dispersion relations under the dipole transformation along \(\hat{\beta}\) (continued).}
    \label{fig:Dipole_Beta_Continued}
\end{figure}

\begin{figure}[ht!]
    \centering
    \begin{subfigure}{0.47\textwidth}
        \includegraphics[width=\textwidth]{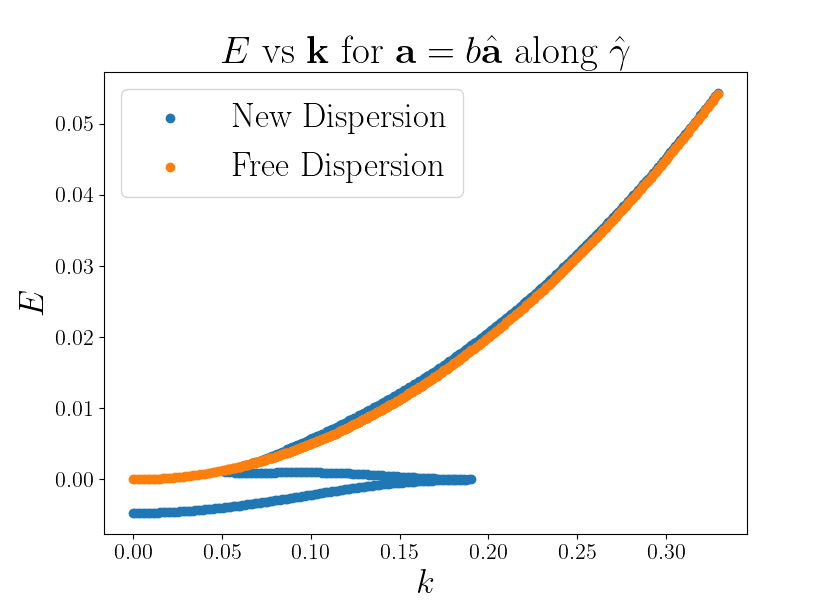}
        \caption{Dispersion relation for \( \mathbf{a} = b\hat{\mathbf{a}} \).}
    \end{subfigure}
    \hfill
    \begin{subfigure}{0.47\textwidth}
        \includegraphics[width=\textwidth]{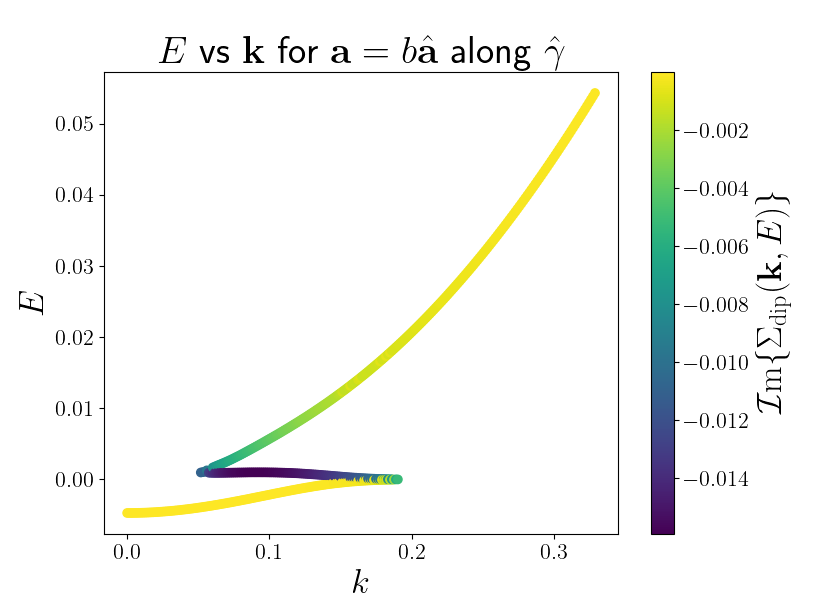}
        \caption{Colormap of the corresponding imaginary part.}
    \end{subfigure}
    \begin{subfigure}{0.47\textwidth}
        \includegraphics[width=\textwidth]{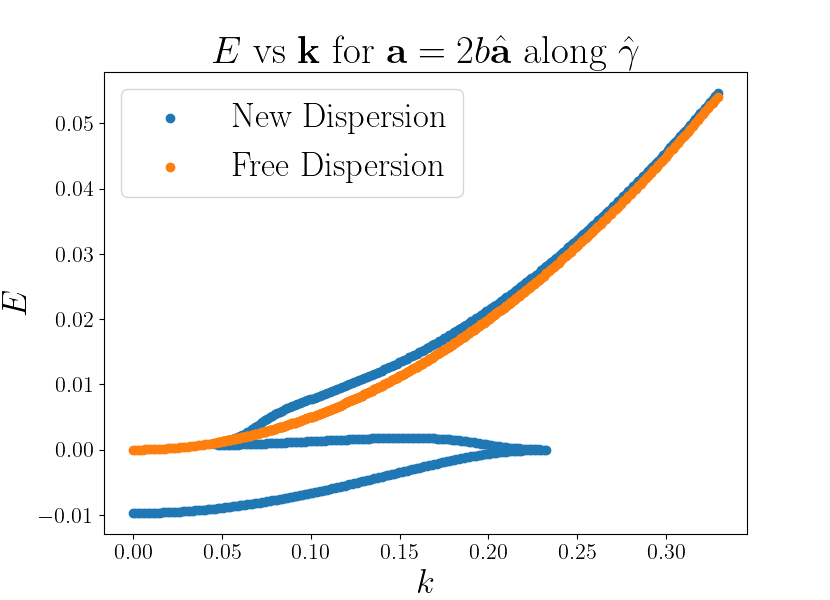}
        \caption{Dispersion relation for \( \mathbf{a} = 2b\hat{\mathbf{a}} \).}
    \end{subfigure}
    \hfill
    \begin{subfigure}{0.47\textwidth}
        \includegraphics[width=\textwidth]{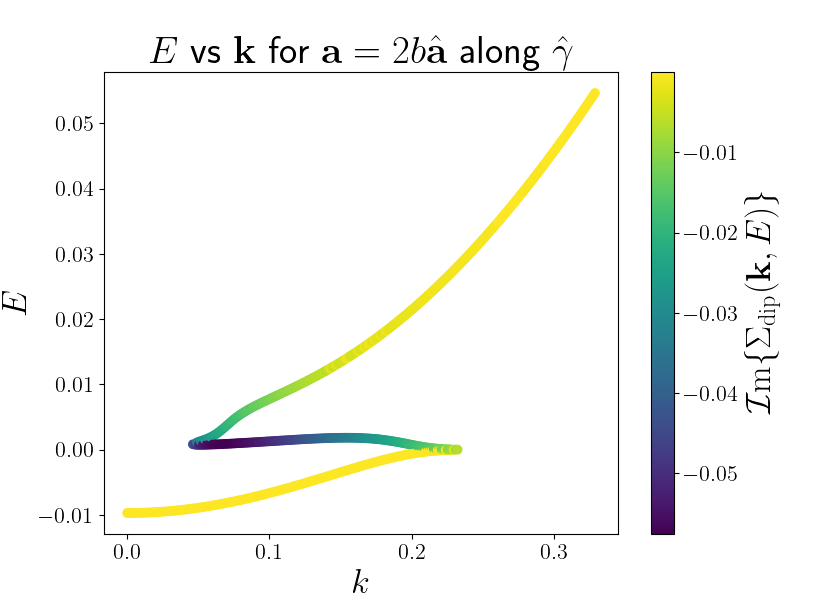}
        \caption{Colormap of the corresponding imaginary part.}
    \end{subfigure}
    \caption{Dispersion relations under the dipole transformation along \(\hat{\gamma}\).}
    \label{fig:Dipole_Gamma}
\end{figure}

\begin{figure}[ht!]
    \centering
    \begin{subfigure}{0.47\textwidth}
        \includegraphics[width=\textwidth]{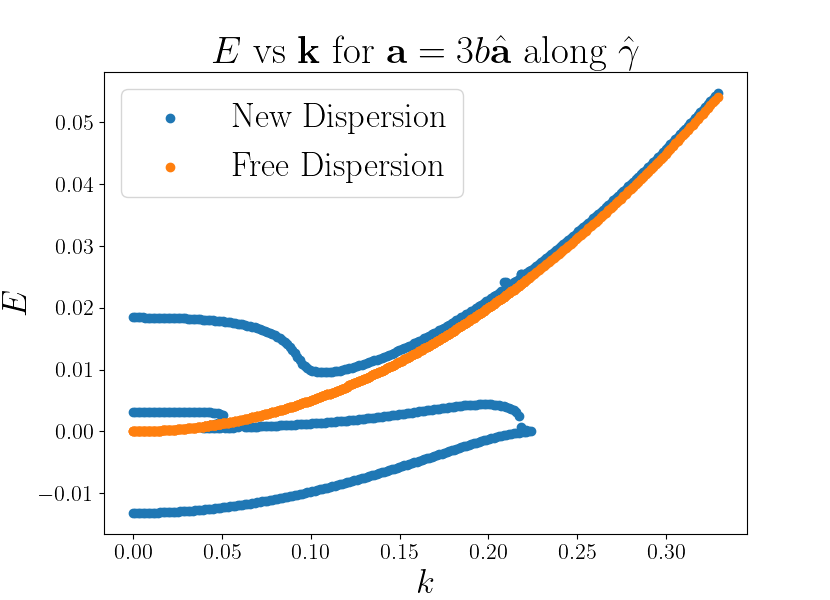}
        \caption{Dispersion relation for \( \mathbf{a} = 3b\hat{\mathbf{a}} \).}
    \end{subfigure}
    \hfill
    \begin{subfigure}{0.47\textwidth}
        \includegraphics[width=\textwidth]{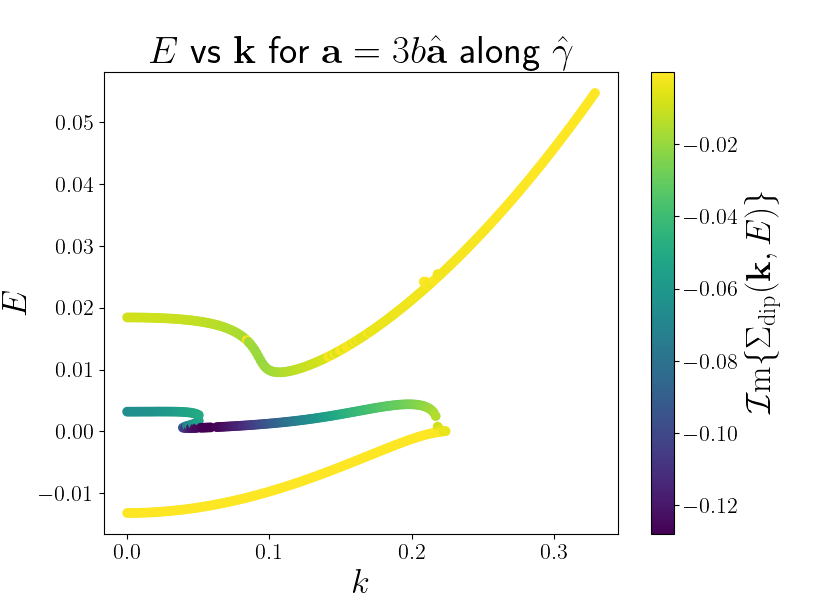}
        \caption{Colormap of the corresponding imaginary part.}
    \end{subfigure}
    \caption{Dispersion relations under the dipole transformation along \(\hat{\gamma}\) (continued).}
    \label{fig:Dipole_Gamma_Continued}
\end{figure}

\begin{figure}[ht!]
    \centering
    \begin{subfigure}{0.47\textwidth}
        \includegraphics[width=\textwidth]{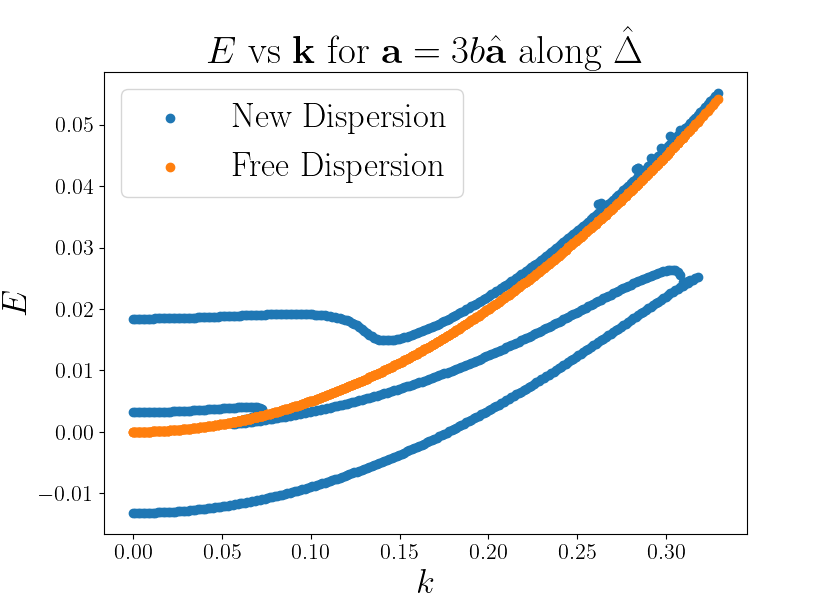}
        \caption{Dispersion relation plotted against the free dispersion curve.}
    \end{subfigure}
    \hfill
    \begin{subfigure}{0.47\textwidth}
        \includegraphics[width=\textwidth]{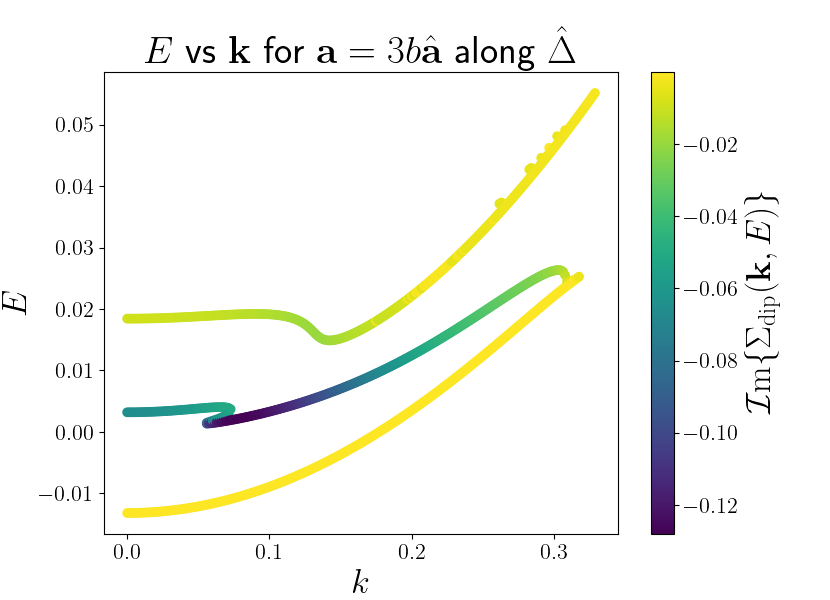}
        \caption{Colormap of the corresponding imaginary part.}
    \end{subfigure}
    \caption{Dispersion relations under the dipole transformation along \(\hat{\Delta}\).}
    \label{fig:Dipole_Delta}
\end{figure}

\begin{figure}[ht!]
    \centering
    \includegraphics[width=0.5\textwidth]{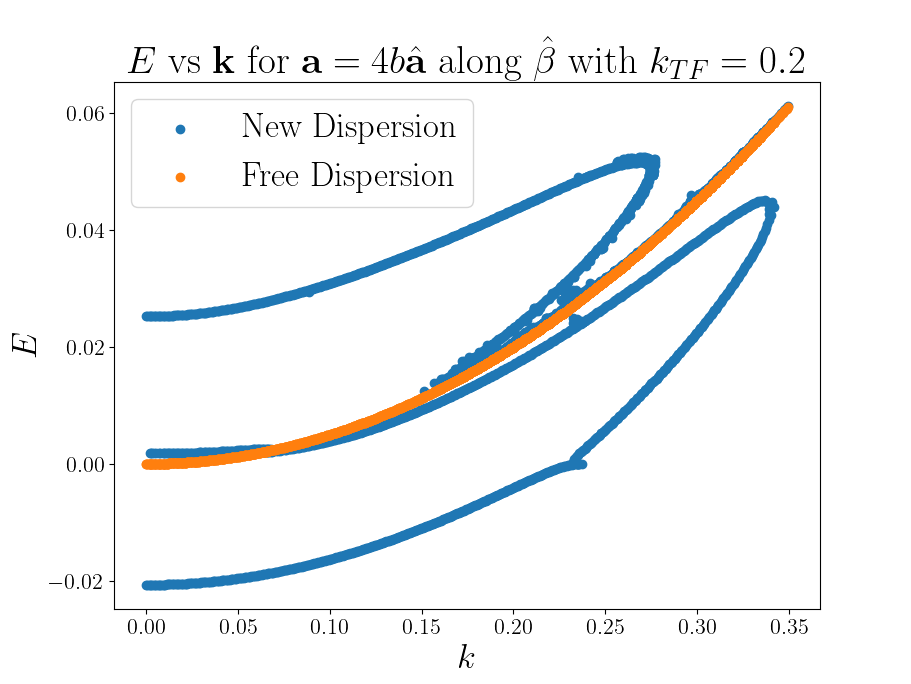}
    \caption{Dispersion relation under the dipole transformation for \( \mathbf{a} = 4b\hat{\mathbf{a}} \) along \(\hat{\beta}\), with \( k_{TF} \) reduced from 0.2336 to 0.2.}
    \label{fig:Dip_4b_along_beta_ktf_0.2}
\end{figure}

\subsection{Dipole Transformation}
The dipole transformation exhibits richer electronic properties. As the strain increases (via increasing the magnitude of $\mathbf{a}$) additional bands emerge, as seen in Figures \ref{fig:Dipole_Alpha} through \ref{fig:Dipole_Gamma}. Notable observations include

\begin{itemize}
    \item Along \(\hat{\alpha}\): All bands are dispersive, as seen in Figures \ref{fig:Dipole_Alpha} and \ref{fig:Dip_4b_along_beta_ktf_0.2}. 
    
    \item Bands with greater overlap with the free dispersion curve at low energies tend to have shorter lifetimes.
    \item Flat Band Formation: Along \(\hat{\beta}\), increasing the strain results in the formation of a flat band, as shown in Figure \ref{fig:Dipole_Beta_Continued}(a) for \(\mathbf{a} = 3b\hat{\mathbf{a}}\). This flat band completely overlaps with the free dispersion curve at high energies. Similar behavior is observed along \(\hat{\gamma}\) (Figure \ref{fig:Dipole_Gamma_Continued}(a)) and \(\hat{\Delta}\) (Figure \ref{fig:Dipole_Delta}).

    \item The Flat Band in Figure \ref{fig:Dipole_Beta_Continued}(a) starts to become dispersive as the strain is further increased by increasing \(\mathbf{a}\) to $4b\hat{\mathbf{a}}$, as seen in \ref{fig:Dipole_Beta_Continued}(c). This becomes more pronounced when the strain is increased further by reducing \(k_{TF}\) from 0.2336 to 0.2, as shown in Figure \ref{fig:Dip_4b_along_beta_ktf_0.2}. These results demonstrate that flat bands do not necessarily form solely due to high strain but emerge under specific strain fields and directions—parameters that can be referred to as ``magic" conditions.
\end{itemize}

Increasing \(k_{TF}\) or decreasing \(\eta_{dis}\), which weakens the electron-dislocation interaction, produces the same trends. Flat bands still form, but with a smaller energy shift relative to the free dispersion curve. These bands form for different strain fields and along different directions. Conversely, if \(k_{TF}\) is made sufficiently large and \(\eta_{dis}\) sufficiently low, the renormalized dispersion curve will completely overlap with the free dispersion curve, as the interaction effect becomes negligible.

It is notable that under the averaged transformation, where the strain is uniform along all directions, no significant changes are observed in the electronic dispersion. This suggests that uniform strain alone does not substantially alter the electronic properties of the system. In contrast, non-uniform strain, as introduced in the dipole transformations, plays a major role in producing interesting electronic properties. It is this non-uniform strain that appears to be responsible for phenomena such as flat band formation, highlighting its critical influence on the system's electronic structure.  Following work in \cite{andrade2023}, which showed that flat bands can be generated in graphene, mapped onto a 1D system, by an oscillating strain with a wavelength that is slightly different than the separation of atoms on the same sublattice, we hypothesize that flat bands form in our model when the average distance between strain modulations (i.e. average distance between high and low regions of strain) is on the order of the average nearest-neighbor distance between dislocation cores with the same Burgers vector on different dislocation dipoles along a given direction. We note here that anisotropic flat bands are observed in our random array of dislocation dipoles, which will have different average strain modulations along different directions. We aim to explore this hypothesis in a more mathematically rigorous way using the statistical schemes we adopt in this paper in future work.

\section{Conclusion}
\label{sec:conclusion}
We have demonstrated that dislocations significantly influence the electronic properties of free carriers in materials. Specifically, we have shown that strain can alter electronic properties and induce anisotropic flat bands along certain directions and under specific strain fields. Our findings emphasize the critical role of non-uniform strain in generating flat bands at these ``magic" parameters.
\newline 
\newline 
\newline
The computational results presented in this work can be reproduced using the provided codebase, available at \cite{azizfallgithub}

\section*{Acknowledgments}
    We thank Ira Rothstein (CMU) for his insightful comments; AFOSR and ARO through the MURI program, NSF (2108784), and the GEM, McGaw and Adamson Fellowships to Aziz Fall for financial support; and NSF for XSEDE computing resources provided by Pittsburgh Supercomputing Center. Aziz Fall thanks Saptarshi Saha (CMU) for visualization of the dislocation displacement field using Ovito \cite{stukowski2010ovito}.

\appendix 

\section{Self-Energy Integral for Random Coulomb Scattering}
\label{appendix}
We showcase the explicit calculation of the integral 
\begin{equation}
    M(E,\mathbf{p},k_{TF}) = \int d^{3}\mathbf{k} \frac{1}{E - \frac{(\mathbf{k} + \mathbf{p})^{2}}{2} + i\epsilon} \frac{1}{(|\mathbf{k}|^{2} + k_{TF}^{2})^{2} }
\end{equation}
We first re-write the integral as 
\begin{equation}
    \frac{\partial}{\partial k_{TF}^{2}} \bigg[ \int d^{3}\mathbf{k} \frac{2}{-2E + (\mathbf{k} + \mathbf{p})^{2} - 2i\epsilon} \frac{1}{(|\mathbf{k}|^{2} + k_{TF}^{2})}    \bigg]
\end{equation} 
Using Feynman parameters the integral turns into 
\begin{equation}
    \frac{\partial}{\partial k_{TF}^{2}} \int_{0}^{1} dx \int d^{3}\mathbf{k} \frac{2}{ ( |\mathbf{k}|^{2} + \chi )^{2}  }
\end{equation}
with 
$\chi \equiv -(x\mathbf{p})^{2} + x|\mathbf{p}|^{2} - 2x(E + i\epsilon) + (1 - x)k_{TF}^2$.
We can now easily compute the integral over $\mathbf{k}$ and $x$ in order to get 
\begin{equation}
    \frac{4 \pi^2}{ |\mathbf{p}|} \frac{\partial}{\partial k_{TF}^{2}} \bigg[ \arctan\bigg( \frac{|\mathbf{p}|}{\sqrt{-2(E + i\epsilon)} - k_{TF}} \bigg)  + \arctan\bigg( \frac{2k_{TF}|\mathbf{p}| }{2(E + i\epsilon) + k_{TF}^2 - |\mathbf{p}|^2 }  \bigg)   \bigg]
\end{equation}
After differentiating with respect $k_{TF}^2$ we get our final result of  
\begin{equation}
    M(E,\mathbf{p},k_{TF}) = \frac{-4\pi^{2}}{2k_{TF} \left( 2k_{TF} \sqrt{-2(E + i\epsilon)} -2(E + i\epsilon) + |\mathbf{p}|^{2} + k_{TF}^{2} \right) }
\end{equation}

\end{document}